\documentclass[journal=jacsat,manuscript=article]{achemso}

\usepackage[version=3]{mhchem} 

\usepackage{siunitx}
\usepackage{xspace}
\usepackage{amssymb}
\usepackage[nohyperlinks,printonlyused,nolist]{acronym}
\usepackage{lineno}

\newcommand*{\nccop}{\ce{NCCO+\xspace}}
\newcommand*{\nccn}{\ce{NCCN\xspace}}

\newcommand*{\NN}{\ce{N$_2$}}
\newcommand*{\wn}{cm$^{-1}$\xspace}

\author{Marcel Bast}
\author{Julian B\"oing}
\author{Thomas Salomon}
\author{Eline Plaar}
\affiliation[PH1]
{I. Physikalisches Institut, Universit\"at zu K\"oln, Z\"ulpicher Str. 77, 50937 K\"oln, Germany.}
\author{Igor Savi\'{c}}
\affiliation[NoviSad]
{University of Novi Sad, Faculty of Sciences, Department of Physics, Novi Sad 21000, Serbia}
\author{Mathias Sch\"afer}
\affiliation[Chemie]
{Institute of Organic Chemistry, Department of Chemistry, University of Cologne, Greinstrasse 4, 50937 K\"oln, Germany}
\author{Oskar Asvany}
\author{Stephan Schlemmer}
\author{Sven Thorwirth}
\affiliation[PH1]
{I. Physikalisches Institut, Universit\"at zu K\"oln, Z\"ulpicher Str. 77, 50937 K\"oln, Germany.}
\email{sthorwirth@ph1.uni-koeln.de}

\title[An \textsf{achemso} demo]
{Spectroscopic detection and characterization of cyanooxomethylium, \nccop}

\keywords{Ion trap, Action spectroscopy, Rotational spectroscopy, Ro-vibrational spectroscopy, Infrared spectroscopy, \nccop, Quantum-chemical calculations}

\begin{document}

\begin{tocentry}
\includegraphics{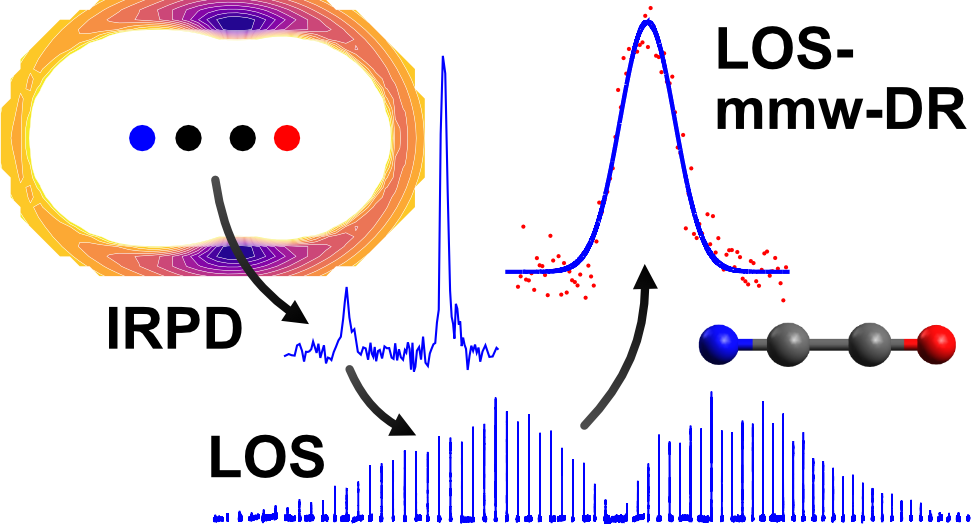}
\end{tocentry}

\begin{abstract}

Cyanooxomethylium, \nccop, a fundamental linear acylium ion, has been observed spectroscopically
for the first time using action spectroscopy in ion trap apparatuses.
A first low-resolution infrared spectrum
was obtained between 500 to 1400\,\wn\ and 2000 to 2500\,\wn\ using the \ac{FELIX} and the FELion apparatus, employing infrared predissociation of the weakly bound \nccop --Ne complex. Subsequently, high-resolution studies of the bare ion were performed with the COLtrap~II setup, one targeted at the CN-stretching mode $\nu_2$ around 2150\,\wn\
using leak-out spectroscopy and one at the pure rotational spectrum employing a leak-out infrared/millimeter-wave double resonance approach covering transition frequencies as high as 246\,GHz. Spectroscopic detection and analysis were guided by high-level quantum-chemical calculations performed at the CCSD(T) level of theory. The collected data permit accurate frequency predictions to support future astronomical searches with sensitive radio telescopes.
\end{abstract}


\begin{acronym}[FWHM]\itemsep0pt
\acro{AMC}{amplifier-multiplier chain}
\acro{CFOUR}{Coupled-Cluster techniques for Computational Chemistry}
\acro{DR}{double resonance}
\acro{FELIX}{Free Electron Laser for Infrared eXperiments}
\acro{FWHM}[FWHM]{Full Width at Half Maximum}
\acro{IR}{infrared}
\acro{IRPD}{infrared predissociation}
\acro{LO}{leak-out}
\acro{LOS}{leak-out spectroscopy}
\acro{mmw}{millimeter-wave}
\acro{MW}{microwave}
\acro{QCL}{quantum cascade laser}
\acro{RF}{radio frequency}
\acro{RMS}{root mean square}
\acro{SIS}{storage ion source}
\acro{SNR}{signal-to-noise ratio}
\end{acronym}

\section{Introduction}

More than 350 molecules have been detected in space to date and their number
continues to grow\footnote{An up-to-date list of astronomically detected molecules with complementary information is maintained at the Cologne Database for Molecular Spectroscopy (CDMS)\cite{mueller_cdms} accessible online at https://cdms.astro.uni-koeln.de/)}. 
About 50 of those species are molecular ions that are both positively and negatively charged. 
A particularly attractive group of ions that has recently triggered astrochemical interest is the group of so-called acylium ions, positively charged species of the general formula $R-$\ce{CO+}. 
The prototypical formylium ion, \ce{H-CO+} (aka protonated carbon monoxide) 
has been known in space for more than 50 years\cite{buhl_nature_228_267_1970, woods_PRL_35_1269_1975} and also was the first polyatomic ion
to be detected there. However, it was not until very recently that two other acylium ions were observed
by their rotational spectra toward Taurus molecular cloud 1 (TMC-1) in the radio regime: linear H$-$C$\equiv$C$-$C$\equiv$\ce{O+}\cite{cernicharo_AAA_642_L17_2020}, as well as CH$_3-$C$\equiv$\ce{O+}\cite{cernicharo_CH3CO+_AA_2021}, a prolate symmetric rotor. To date, these three
species also remain the only acylium ions studied at high spectral resolution
in the laboratory. \cite{cernicharo_AAA_642_L17_2020,cernicharo_CH3CO+_AA_2021,asvany_hc3op_PCCP_2023}.
Consequently, high-resolution spectroscopic studies of other simple acylium ions are very desirable to check for their astronomical presence in molecular clouds and to evaluate the astrochemical relevance of acylium ions from a more general perspective.

In the present paper, both low- and high-resolution spectroscopic investigations of a hitherto spectroscopically not known acylium ion will be presented, linear tetratomic cyanooxomethylium, N$\equiv$C$-$C$\equiv$\ce{O+}. Despite a number of theoretical studies 
performed over the years\cite{pykko1991_JMStTheo_234_269_1991,mcgibbon_IJMSIP_121_R11_1992,francisco_JCP_107_3840_1997,jursic_JMStTheo_460_207_1999,yu_JPCA_109_2364_2005,chi_JMStTheo_763_91_2006} the only experimental characterization of \nccop\ appears to be that of McGibbon et al. reporting on the presence of \nccop\ in
mass spectra of pyruvonitrile (acetyl cyanide, \ce{CH3C(O)CN}) and methyl cyanoformate (\ce{CH3OC(O)CN}).\cite{mcgibbon_IJMSIP_121_R11_1992}
In the present investigation, selected spectroscopic properties of \nccop\ have been studied in some detail. Low-resolution \ac{IR} spectra were obtained via predissociation of its weakly bound complex with Ne followed by high-resolution
\acf{LOS} of the $\nu_2$ vibrational fundamental 
at 2150\,\wn . Finally, the pure rotational spectrum was observed employing a \ac{LO} infrared/millimeter-wave \ac{DR} scheme.
The spectroscopic findings and analyses, along with results from complementary high-level quantum-chemical calculations, will be summarized in the following.

\section{Experimental methods}

Spectroscopic studies of \nccop\ were performed over the course of 
three consecutive measurement campaigns using two of our cryogenically cooled 22-pole ion trap apparatus. 
Experimental setups and the applied action spectroscopic methods are described in some detail in the following.

The first spectroscopic detection of \nccop\ ions 
was achieved using the cryogenic 4\,K 22-pole ion trap apparatus FELion in combination with the \acf{FELIX} at Radboud University (Nijmegen, The Netherlands)~\cite{oepts_IPT_36_297_1995}.
The experimental setup and typical experimental conditions have been described in detail elsewhere previously\cite{jusko_FD_2019}.
Briefly, \nccop\ ($m/z~=~54$) was generated in a \ac{RF} \ac{SIS} by electron bombardment of 
commercially available 
pyruvonitrile (\ce{CH3C(O)CN}) using an electron energy of about 55\,eV. 
After mass selecting the pulsed ion beam for $m/z~=~54$ in the first quadrupole mass filter (QP~I), the ions were 
trapped
in the 22-pole ion trap~\cite{asv10} 
and cooled down to a nominal temperature of about 7\,K using a short pulse of a 3:1 mixture of He:Ne. 
In such an environment, kinetically and internally cooled ions can form weakly bound ion--Ne complexes 
through three-body collisions with the noble gas atoms. After trapping times of 1 to 4\,s, 
the trap content was analyzed with a second quadrupole (QP~II) and a Daly-type detector. 
Initially, more than 100,000 \nccop\ ions were trapped per measurement cycle 
and found to form \nccop --Ne complexes ($m/z~=~74$) with an efficiency of about 20\%.
A corresponding mass spectrum is shown in the Supporting Information (Figs.~S1 - S2). 
To record \ac{IRPD} spectra of the \nccop --Ne weakly bound complex,
the ion trap content was irradiated with pulsed \ac{IR} radiation of the free electron laser at a repetition rate of 10\,Hz and with energies of up to a few tens of mJ per pulse. 
For the \ac{IRPD} scheme used here, 
the number of singly tagged ions ($m/z$~=~74, mass selected via QP~II) is continuously monitored while \ac{FELIX} is tuned.
In case the photon energy matches the excitation energy of a vibrational mode of the cluster, 
the latter dissociates, which leads to the \ac{IRPD} spectrum, recorded as a depletion of the ion signal.

\begin{figure}[h]
\centering
  \includegraphics[width=\linewidth]{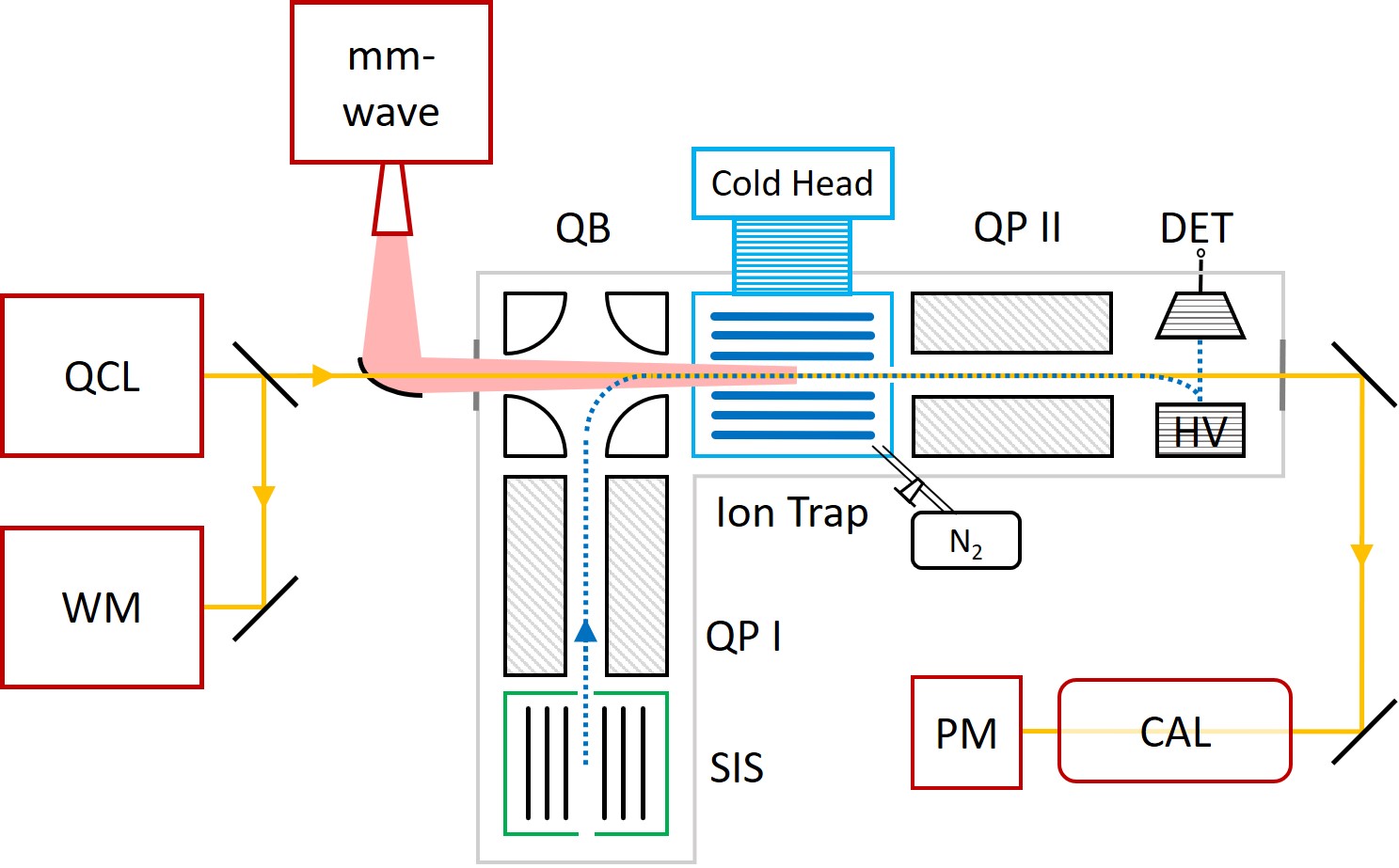}
  \caption{
  Schematic view of the 10 K 22-pole ion trap apparatus COLtrap II (adapted from Figure 1 in Ref.\cite{bast_JMS_398_111840_2023}\,; 
  reproduced with permission from Elsevier). The setup is extended here by a millimeter-wave (mmw) setup for IR/mmw double resonance experiments, consisting of a synthesizer, an amplifier-multiplier chain and a horn antenna. The mmw beam (shown in pale red) is focused by an elliptical mirror and enters the vacuum chamber through a diamond window. A small hole in this mirror also allows the IR laser (shown in yellow) to pass along the axis. See the text for further details.
  }
  \label{fig:COLtrap_II_Scheme_IR_MW}
\end{figure}

High-resolution observations of rotational-vibrational spectra were subsequently 
performed using the
COLtrap~II ion trap apparatus~\cite{bast_JMS_398_111840_2023},
which was recently upgraded to also enable millimeter-wave measurements
(Fig.~\ref{fig:COLtrap_II_Scheme_IR_MW}).
These measurements were performed using \ac{LOS}~\cite{schmid_JPCA_126_8111_2022}, 
allowing to record IR spectra of bare ions rather than those of weakly bound clusters.
In this action spectroscopy scheme, the internal vibrational energy of excited 
ions is partially converted into kinetic energy of the ions and a neutral collision partner (here \ce{N2}), 
allowing the target ion to overcome the barrier of the exit potential of the ion trap
and be detected.
In these high-resolution experiments, \nccop\ ions were produced 
from methyl cyanoformate (\ce{CH3OC(O)CN}) 
using an electron energy of about 65\,eV
(for a mass spectrum see Figs.~S3 - S4),
mass selected in QP~I (see Fig.~\ref{fig:COLtrap_II_Scheme_IR_MW}) and
injected into the 22-pole ion trap. 
The nominal trap temperature was kept at 42(2)\,K to 
avoid freeze-out of 
\NN, which was 
continuously introduced into the trap at a number density on the
order of $10^{12}$\,cm$^{-3}$. An additional initial helium pulse ($\sim 10^{15}$\,cm$^{-3}$) was used for efficient trapping and cooling.
High-resolution leak-out spectroscopy was performed around 2150\,\wn\ in the CN-stretching region of \nccop\ with
a narrow line width quantum cascade laser (QCL, Daylight Solutions MHF 1961 - 2205\,\wn, 200\,mW) 
using an irradiation time of about 400\,ms per 
trapping cycle.
For wavenumber calibration, a small portion (5\%) of the infrared radiation is coupled into a wavemeter (WM, 
Bristol Instruments, 771A-MIR, quoted accuracy $10^{-3}$\,\wn )
via a CaF$_2$ window acting as a beam splitter.
In the spectroscopy scheme used here, 
the number of ions leaking out of the trap was counted during the trapping time.
Along the way to the detector, the ions pass through QP~II, which selects ions with $m/z~=~54$. 
An off-resonance background LO rate of about 1\% per cycle of the total trap content 
(around 40,000 \nccop~counts per cycle) turned out to provide a high S/N ratio. 
This procedure results in almost background free ion signals upon excitation of a rotational-vibrational transition. 
To account for potential drifts of experimental settings during long-time measurements, 
the wavenumber ($\Tilde\nu$) dependent experimental counts were normalized according to 
the background LO rate (BG): 
$\rm{norm.~signal~(\Tilde\nu)} = \frac{\rm{Counts}~(\Tilde\nu) - \rm{BG}}{\rm{BG}}$ (see, Ref. \citenum{bast_JMS_398_111840_2023}).

As the photon energy of a pure rotational transition is too low to drive the LO
process, an \ac{IR}/mmw \ac{DR} scheme~\cite{asvany_hc3op_PCCP_2023} 
has been applied in a third measurement campaign to record rotational transitions of \nccop.  
In this approach, the infrared radiation that drives a selected rotational-vibrational transition is kept fixed on-resonance.
Meanwhile, the \ac{mmw} radiation is tuned in fixed steps of 5 to 10 kHz with the corresponding LO rate being monitored.
This allows to search for rotational transitions sharing an energy level with the lower rotational level of the infrared transition.
The required microwave radiation is provided by a synthesizer (Rohde \& Schwarz SMF 100A) referenced to a 10\,MHz rubidium atomic clock
and driving an amplifier-multiplier chain (AMC, Virgina Diodes Inc.\ WR 9.0) 
which upconverts (x9) it into the millimeter-wave range. An additional frequency doubler is used to extend the accessible
frequency range even further, providing frequencies as high as 250\,GHz.
Focused by a horn antenna and an elliptical mirror ($f = 43.7$ mm \cite{jusko_PRL_112_253005_2014}), 
the \ac{mmw} radiation enters the vacuum chamber through a diamond window. 
The mirror features a central hole through which the IR beam from the QCL can pass and overlap 
with the mmw-radiation in the ion trap as needed for DR spectroscopy.


\section{Quantum-chemical methods}
\label{QCCsection}

Several computational studies of \nccop\ performed at various (lower) levels of theory have been reported
in the literature previously, primarily aimed at the calculation of structural parameters, harmonic force fields
and energy ordering of structural isomers within the [2C,N,O\ce{]+}
family\cite{pykko1991_JMStTheo_234_269_1991,mcgibbon_IJMSIP_121_R11_1992,francisco_JCP_107_3840_1997,jursic_JMStTheo_460_207_1999,yu_JPCA_109_2364_2005,chi_JMStTheo_763_91_2006}.
In the present study, quantum-chemical calculations have been performed
at the coupled-cluster singles and doubles (CCSD) level 
augmented by a perturbative treatment of triple excitations, 
CCSD(T)\cite{raghavachari_chemphyslett_157_479_1989}, 
together with correlation consistent (augmented) polarized valence 
and (augmented) polarized weighted core-valence basis sets,
i.e., cc-pV$X$Z, \cite{dunning_JCP_90_1007_1989} 
aug-cc-pV$X$Z,\cite{dunning_JCP_90_1007_1989, kendall_JCP_96_6796_1992,woon_JCP_98_1358_1993} 
cc-pwCV$X$Z, \cite{dunning_JCP_90_1007_1989,peterson_JCP_117_10548_2002} 
and aug-cc-pwCV$X$Z \cite{dunning_JCP_90_1007_1989, kendall_JCP_96_6796_1992,peterson_JCP_117_10548_2002}
(with $X$=T, Q)
as well as atomic natural orbital basis sets ANO$X$ ($X$=1,2) \cite{almlof_JCP_86_4070_1987}.
Equilibrium geometries have been calculated using analytic gradient techniques
\cite{watts_chemphyslett_200_1-2_1_1992}, while
harmonic frequencies have been computed using analytic second-derivative techniques
\cite{gauss_chemphyslett_276_70_1997,stanton_IntRevPhysChem_19_61_2000}.
For anharmonic computations second-order vibrational perturbation theory (VPT2) \cite{mills_alphas}   
has been employed and additional numerical differentiation of analytic second derivatives has been applied 
to obtain the third and fourth derivatives required for the application of VPT2 \cite{stanton_IntRevPhysChem_19_61_2000,stanton_JCP_108_7190_1998}.
Usage of the frozen core approximation has been indicated throughout by ``fc''
while ``ae'' indicates usage of all electrons in the correlation treatment.
All calculations have been performed using 
the CFOUR program package \cite{cfour,harding_JChemTheoryComput_4_64_2008}.

To further improve calculated spectroscopically relevant  molecular parameters, scaling factors
$X_{exp}/X_{calc}$ (with $X$ being a parameter of choice, e.g., $B$, $D$, 
$eQq$, $\tilde{\nu}$) may be derived 
using isoelectronic and isostructural species that are known from experiment (see, e.g., Refs.\citenum{martinez_JCP_138_094316_2013,thorwirth_MolPhys_118_e1776409_2020,cernicharo_AAA_642_L17_2020,cabezas_AAA_659_L8_2022}). 
For \nccop\ as a 26 electron system, a sizable number of potentially applicable molecular ``calibrators''
are available from previous high-resolution spectroscopic studies to the extent
that their ground state rotational constant
$B_0$ and centrifugal distortion parameter $D_0$ are known to high accuracy:
NCCN\cite{maki_JMS_269_166_2011}, CNCN\cite{gerry_JMS_140_147_1990}, \ce{C3O}\cite{bizzocchi_AaA_492_875_2008}, \ce{C3N^-}\cite{amano_JMS_259_16_2010}, \ce{C4H^-} \cite{amano_JCP_129_244305_2008},
\ce{HC3N}\cite{thorwirth_JMolSpectrosc_204_133_2000}, HCCNC\cite{guarnieri_JMS_156_39_1992}, \ce{HC3O^+}\cite{asvany_hc3op_PCCP_2023}, \ce{NCCNH^+}\cite{gottlieb_JCP_113_1910_2000}, \ce{HC3NH^+}\cite{gottlieb_JCP_113_1910_2000}, and \ce{HCCNCH^+}\cite{agundez_AAA_659_L9_2022}.
Beyond parameters relevant to the ground vibrational state, cyanogen, NCCN, has been studied in the infrared comprehensively\cite{maki_JMS_269_166_2011} such as to serve
as a useful calibrator for the vibrational spectrum of \nccop .

Structural information about the weakly bound cluster of \nccop\ with Ne was derived at the
fc-CCSD(T)/aug-cc-pVTZ level of theory using a strategy applied recently on similar
complexes of other fundamental molecular ions\cite{thorwirth_MolPhys_118_e1776409_2020,thorwirth_mol_29_665_2024}.


\section{Results and discussion}

\subsection{Structures of \nccop\ and its weakly bound complex with Ne}

The equilibrium structural parameters of \nccop\ calculated at the CCSD(T) level of theory using a variety of basis sets are collected in
Table S1 (see the Supporting Information). At the ae-CCSD(T)/cc-pwCVQZ level of theory, that is known to provide
very high quality equilibrium structural parameters for molecules
containing first- and second-row atoms\cite{bak_JCP_114_6548_2001,coriani_JCP_123_184107_2005}, the bond lengths
are $r_{\rm N-C}=1.1668$\,\AA , $r_{\rm C-C}=1.3692$\,\AA , and $r_{\rm C-O}=1.1163$\,\AA\ and the center-of-mass frame dipole moment is 1.62\, D.

\begin{figure}[h!] 
\centering
    \includegraphics[width=\linewidth]{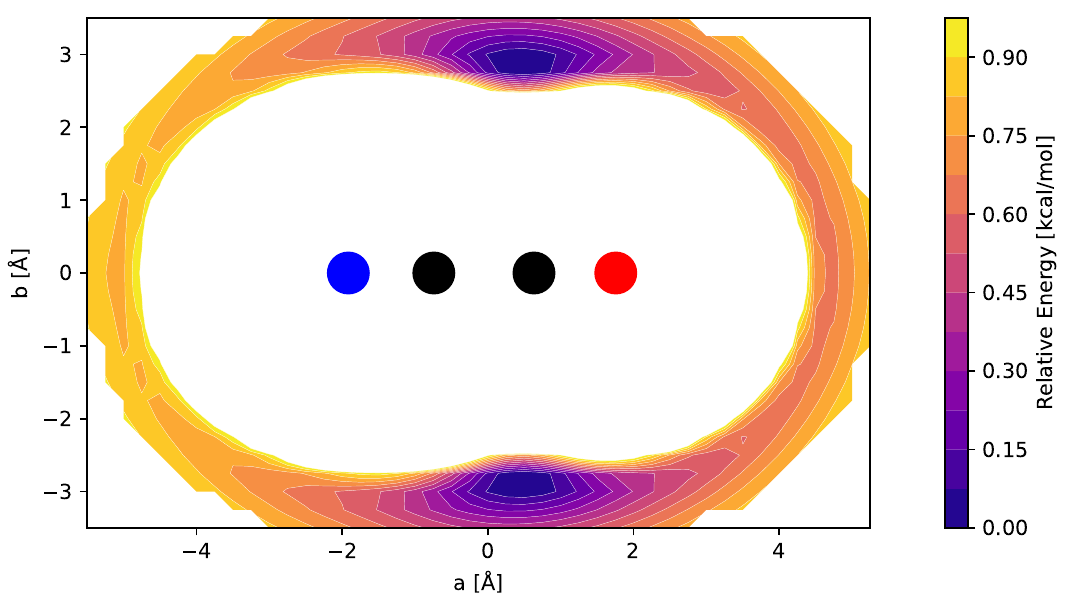}
    \caption{Contour plot of \nccop --Ne showing the fc-CCSD(T)/aug-cc-pVTZ potential energy as a function
    of the Ne atom position covering the interval [0.075, 0.975] kcal/mol in
    steps of 0.075 kcal/mol above the global minimum. Atom color code: nitrogen (blue), carbon (black), and oxygen (red). 
    }
    \label{fig:Potentialoberflaeche_NCCO+_Ne}
\end{figure}

The energetically favorable position(s) of Ne with respect to \nccop\ were determined through calculation of a potential energy map. To do this, the
fc-CCSD(T)/aug-cc-pVTZ equilibrium structure of \nccop\ was kept fixed while the position of the Ne atom was used to sample a 10$\times$6 \AA $^2$ grid at a spacing of 0.25\,\AA\ and distances ranging from 1.5 to 3.5\,\AA\ about the ion. The potential energy map is shown in Fig. \ref{fig:Potentialoberflaeche_NCCO+_Ne}
and reveals the global minimum to assume a T-shaped structure. Possibly, a very shallow minimum is observed for a linear arrangement in which Ne is attached to N.
For advanced structural refinement, the T-shaped global minimum structure was fully optimized at the fc-CCSD(T)/aug-cc-pVTZ level of theory. As with other Ne-tagged weakly bound clusters,
the impact of Ne when binding to \nccop\ is small: Bond lengths between the tagged and bare variants agree to within 2$\times 10^{-4}$\,\AA\ and the deviation to linearity in the cluster does not exceed 0.5$^\circ$ (see Supporting Information). 
As a consequence of this, Ne-tagging is not expected to influence the vibrational spectrum of \nccop\
by much. Indeed, from fc-CCSD(T)/aug-cc-pVTZ harmonic force field calculations the tagging-induced shifts are found to not exceed some 3\,\wn\ (Table \ref{tab:influence-Ne-tagging}). At the same theoretical level, an estimate of the bond dissociation energy $D_0$ yields a value of 1.0 kcal/mol.

\begin{table}[H]
\caption{Harmonic vibrational wavenumbers $\omega_i$ of bare \nccop\ and its T-shaped weakly bound complex with Ne evaluated at the fc-CCSD(T)/aug-cc-pVTZ level of theory (in \wn).}
\label{tab:influence-Ne-tagging}
  \begin{tabular*}{0.48\textwidth}{@{\extracolsep{\fill}}lrr}
  \hline
Mode\textsuperscript{a} & \nccop   &  \nccop --Ne \\
\hline 
$\omega_1$                       & 2376 &  2377 \\
$\omega_2$                       & 2168 &  2169 \\
$\omega_3$                       &  856 &   856 \\
$\omega_4$\textsuperscript{b,c}  &  529 &  527/529  \\
$\omega_5$\textsuperscript{b,c}  &  178 &  179/181  \\
$\omega_6$\textsuperscript{d}    & --   &   72  \\
$\omega_7$\textsuperscript{d}    & --   &   37  \\ \hline
\end{tabular*}\\
$^a$ Mode index borrowed from untagged \nccop~for the sake of comparability.\\
$^b$ Doubly degenerate bending mode in linear species.\\
$^c$ Degeneracy is lifted in the Ne-tagged T-shaped species.\\
$^d$ Extra low-frequency vibrational modes introduced in the \nccop --Ne complex, arbitrary mode index.
\end{table}

\subsection{Broadband \ac{IRPD} spectrum of \nccop --Ne} 

\begin{figure*}[h]
\centering
  \includegraphics[width=\linewidth]{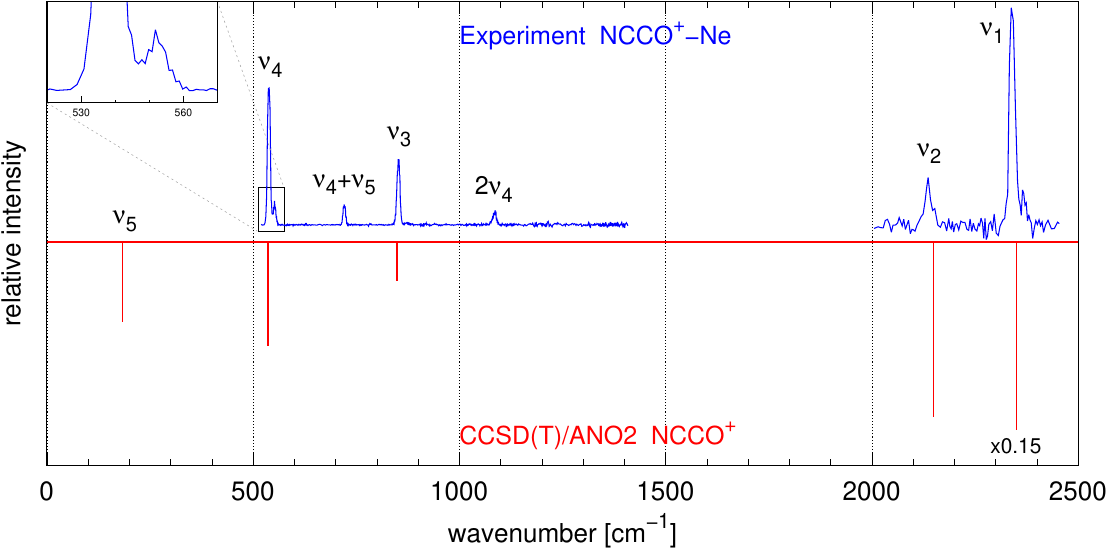}
  \caption{\ac{FELIX} \ac{IRPD} spectrum of \nccop --Ne (blue trace, top) obtained using the FELion ion trap instrument at a nominal trap temperature of 7\,K. 
  Location of the fundamental vibrational bands calculated at the fc-CCSD(T)/ANO2 level of theory is shown as inverted red sticks for comparison.
  }
  \label{fig:FELIX}
\end{figure*}

The \ac{FELIX} \ac{IRPD} spectrum of \nccop --Ne obtained with the FEL\-ion apparatus is shown in Fig.~\ref{fig:FELIX}.
During these experiments, two broad regions were covered, one between 500 to 1400\,\wn\ and another one from 2000 to 2450\,\wn . As a tetratomic linear molecule, \nccop\ possesses five
vibrational fundamentals, three stretching modes and two doubly
degenerate bending modes (Table \ref{tab: Vib NCCN NCCO+ Exp Calc}). 
The IRPD spectrum exhibits six features, of which the four most intense can be assigned readily to vibrational fundamentals of \nccop , or more precisely, of NCCO–Ne: the $\nu_{1}$-, $\nu_{2}$-, and $\nu_{3}$-stretching modes and the $\nu_{4}$-bending mode, in agreement with anharmonic force field calculations of bare \nccop\ at the fc-CCSD(T)/ANO2 level (Table \ref{tab: Vib NCCN NCCO+ Exp Calc}).
The energetically lowest bending mode $\nu_5$ calculated
at 184\,\wn\ was not covered spectroscopically, as were not any fundamental modes related to
the Ne-tag which would occur at even lower wavenumbers (Table \ref{tab:influence-Ne-tagging}). The $\nu_4$ mode at 538\,\wn , however, shows a weak blue-shifted satellite band at 552\,\wn\ that might be a combination mode under participation of Ne; similar features have been observed previously in \ac{IRPD} spectra of other tagged ions
\cite{Pivonka_JCP_118_5275_2003,brunken_JPCA_123_8053_2019,thorwirth_mol_29_665_2024}. The $\nu_1$ mode of \nccop\ might also show such a `tag-satellite'.

\begin{table*}[h]
\footnotesize
  \caption{Fundamental vibrational wavenumbers $\tilde\nu_i$ of NCCN and \nccop\ (in \wn ) and IR band
intensities of \nccop\ (km/mol) as calculated at the fc-CCSD(T)/ANO2 level of theory.
}
  \label{tab: Vib NCCN NCCO+ Exp Calc}
  \begin{tabular*}{1\textwidth}{@{\extracolsep{\fill}}lrrrrrrrrrr}
    \hline
    & \multicolumn{3}{c}{\nccn} & \multicolumn{6}{c}{\nccop} \\ \cline{2-4} \cline{5-11}
    Mode$^a$ & Harm & Anharm & Exp$^b$ & Mode & Harm & Anharm & BE$^c$ & Exp$^d$ & \acs{FWHM} & Int \\
    \hline
        $\nu_1(\sigma_g^+)$ & 2373 & 2333  & 2330.49 & $\nu_1(\sigma)$            & 2391 & 2351 & 2348 & 2340(1)      & 16 & 440  \\
    $\nu_3(\sigma_u^+)$ & 2186 & 2155  & 2157.82 & $\nu_2(\sigma)$                & 2182 & 2149 & 2152 & 2137(1)      & 20 &  68  \\
    $\nu_2(\sigma_g^+)$ &  860 &  840  &  845.59 & $\nu_3(\sigma)$                &  861 &  849 &  855 &  852(1)      & 8  &  22  \\
    $\nu_4(\pi_g)$      &  504 &  502  &  502.77 & $\nu_4(\pi)$                   &  538 &  536 &  537 &  538(1)      & 8 &  39  \\
    $\nu_5(\pi_u)$      &  235 &  233  &  233.72 & $\nu_5(\pi)$                   &  182 &  184 &  185 &  183(4)$^*$ & &  30  \\ \hline
                        &   &       &        & $2\nu_4(\sigma,~\Delta)$           &      &      &      & 1085(1)      & 11 &   \\
                        &   &      &         & $\nu_4+\nu_5(\sigma,~\Delta)$      &      &      &      &  721(1)      & 7 &   \\
    \hline
  \end{tabular*}
$^a$ Ordering of $\nu_2$ and $\nu_3$ bands reversed to match $\nu_i$ ordering of \nccop\ ($C_{\infty v}$).\\ 
$^b$ Experimental vibrational wavenumbers from Ref. \cite{maki_JMS_269_166_2011}. \\
$^c$ Best estimate (BE) values derived from using NCCN as calibrator, see text for details. \\
$^d$ Experimental vibrational wavenumbers (and uncertainties) from \ac{IRPD} of \nccop --Ne complex as derived from Gaussian profile fitting. Typical wavenumber accuracy of FELIX amounts to about 0.5\%. \\
$^*$ Estimation from combination band $\nu_4+\nu_5$ and $\nu_4$ fundamental.
\end{table*} 
As can also be seen from Table~\ref{tab: Vib NCCN NCCO+ Exp Calc}, empirical scaling of the fundamental vibrational wavenumbers of \nccop\ using isoelectronic NCCN has a minuscule effect only because the calculated and experimental anharmonic vibrational wavenumbers of the latter are very close. Hence, the best estimate (BE) values of \nccop\ hardly differ from the unscaled fc-CCSD(T)/ANO2 wavenumbers.
The agreement between the experimental and the calculated anharmonic wavenumbers/best estimate (BE) of  
$\tilde\nu_i$ is remarkable (Table \ref{tab: Vib NCCN NCCO+ Exp Calc}) as deviations are well within the experimentally observed \ac{FWHM} of the bands. The largest discrepancy is observed for the $\nu_2$ fundamental but still amounts to some 12 to 15\,\wn\ or 0.6 to 0.7\% only 
which is comparable to the typical FELIX wavenumber accuracy of about 0.5\%.
Actually, in the absence of any high-level force field calculations, spectroscopic assignment of the \nccop\ \ac{IRPD} spectrum would have been feasible rather effortlessly by direct comparison against the NCCN vibrational wavenumbers alone \cite{maki_JMS_269_166_2011} taking into account that in
a $C_{\infty v}$ species such as
\nccop\ all vibrational modes are infrared active.

Finally, the two weaker features at 721\,\wn\ and 1085\,\wn\ are assigned to the $\nu_{4}+\nu_{5}$ combination band and the $2\nu_{4}$ overtone, respectively, in good agreement with anharmonic force field predictions. From the former, the $\nu_{5}$ wavenumber can be estimated by subtraction as $\sim$183\,\wn, in very good agreement with the calculated value (Table \ref{tab: Vib NCCN NCCO+ Exp Calc}).

\subsection{High-resolution leak-out spectrum}

\begin{figure*}[h]
\centering
  \includegraphics[width=\linewidth]{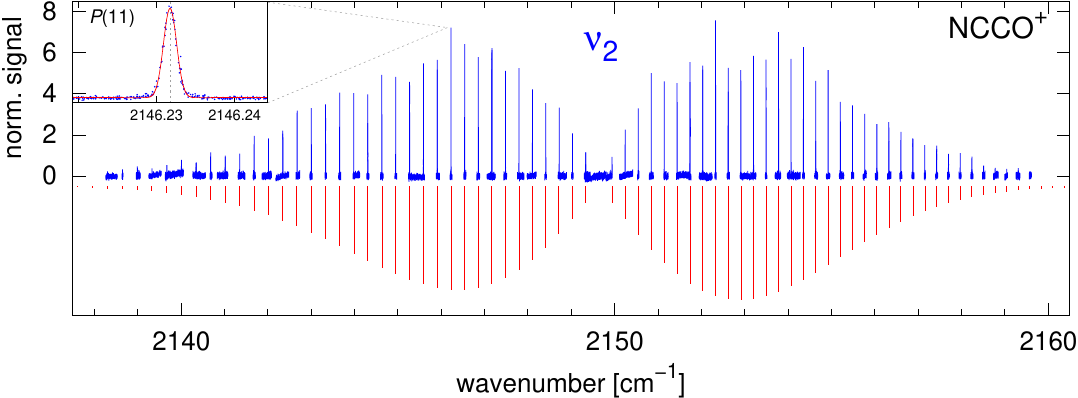}
  \caption{LO spectrum of the $\nu_2$ stretching mode of bare \nccop~(blue trace) recorded at 
  a nominal trap temperature of 42(2)\,K using 10$^{12}$ cm$^{-3}$ of \NN~as a neutral collision partner. 
  Normalized signal is plotted against the wavenumber. 
  A simulation of the rotational-vibrational lines assuming a rotational temperature of 50\,K is shown for comparison as inverted red sticks. 
  An enlarged view of the $P$(11) transition is shown in the top left corner. 
  }
  \label{fig:LOS IR COLtrap II}
\end{figure*}

As a linear species with an ordinary $^1\Sigma$ electronic ground state, at high spectral resolution, any stretching vibrational fundamental 
of \nccop\ will decompose into characteristic $P$- and $R$-branches 
comprising roughly equally spaced rotational-vibrational transitions governed by the $\Delta J=\pm 1$ selection rule. Guided by the experimental low-resolution \ac{IRPD} spectrum 
(Fig. \ref{fig:FELIX})
and
the high-level force field calculations (Table \ref{tbl:Fitparameter_COLtrapII_IR_Exp_und_calc}) the high-resolution IR spectrum of \nccop\ 
was targeted 
using the ion trap apparatus COLtrap~II 
\cite{bast_JMS_398_111840_2023}. 
First lines were detected at about 2140\,\wn\ and spectroscopic assays 
showed these to belong to the $P$-branch tail at comparably high rotational quantum numbers (Fig. \ref{fig:LOS IR COLtrap II}). Scanning upwards, transitions were covered one after another until the 
$\nu_2$ band center
was finally identified at about 2150\,\wn\ in very good agreement 
with the high-level calculation and best-estimate values and blueshifted by about 13\,\wn\ from the IRPD value (Table \ref{tab: Vib NCCN NCCO+ Exp Calc}).
\begin{table*}[h]
  \caption{Calculated, best estimate (BE) and experimental molecular parameters of \nccop~in the ground 
  vibrational and $\nu_2$ excited states (in MHz, unless noted otherwise).}
  \label{tbl:Fitparameter_COLtrapII_IR_Exp_und_calc}
  \begin{tabular*}{1\textwidth}{@{\extracolsep{\fill}}lrrrrr}
    \hline
    Parameter                   & Calc$^a$ & BE$^b$    & IR            &  mmw                   & IR \& mmw        \\ \hline
    $B_e$                       & 4554.302 & $\cdots$  & $\cdots$      & $\cdots$               &  $\cdots$          \\
    $\Delta B_0$                & $-$5.819 & $\cdots$  & $\cdots$      & $\cdots$               &  $\cdots$          \\
    $B_0$                       & 4560.121 &  4564.6   & 4563.775(43)  & 4563.77641(15)         & 4563.77641(15)      \\
    ${D}\times 10^3$$^{,c}$     &    0.537 &  0.592    & 0.603(26)     & 0.60956(14)            & 0.60956(14)        \\ 
    $\alpha_2$                  &   17.002 &  17.07    & 17.0746(19)   &  $\cdots$              & 17.0745(19)          \\
    $\tilde\nu_2$ /\,\wn        & 2149.3   & 2152      & 2149.64318(4) &  $\cdots$              & 2149.64318(4)      \\
    $eQq$(N)                    & $-$5.880 & $-$5.827  &  $\cdots$     &  $\cdots$              & $\cdots$            \\
    $rms_{\rm IR}\times 10^4$ /\,\wn  & $\cdots$ & $\cdots$  &   2.0   &    $\cdots$            &     2.0                 \\
    $rms_{\rm mmw}\times 10^3$             & $\cdots$ & $\cdots$  &  $\cdots$     &   $7.7$                &     7.7                 \\    
    \hline
  \end{tabular*}
  \\
$^a$ $B_e$ and $eQq$(N) calculated at the ae-CCSD(T)/cc-pwCVQZ level; $\Delta B_0$, $D$, $\alpha_2$, and $\tilde{\nu}_2$ calculated at the fc-CCSD(T)/ANO2 level.\\
$^b$ BE values derived from using NCCN as calibrator except for $eQq$(N) that was scaled using \ce{NCCNH+},
see text for details.\\
$^c$ Calculation yields equilibrium value, $D=D_e$; In the IR and global fits, a common constant has been used for both states, $D=D_0~=D_2$.
\end{table*}

Overall, 70 rotational-vibrational lines were finally observed in the wavenumber
range from 2138 to 2160\,\wn\ covering
$P(35)$ to $R(34)$. 
The resulting high-resolution LO spectrum 
is depicted in Fig.~\ref{fig:LOS IR COLtrap II}.
A complete list of the $\nu_2$ transition wavenumbers is provided in the Supporting Information.
Generally, the transitions are found to exhibit Gaussian line shapes with an average FWHM of about 0.002\,\wn , see inset in Fig.~\ref{fig:LOS IR COLtrap II}, and intensities that are well described by an absorption spectrum with a rotational temperature of about 50\,K (cf., Fig. S6). 
This suggests that there is no
influence of the rotational state on the leak-out process.
For a more detailed discussion consult the Supporting Information.

As seen in Fig.~\ref{fig:LOS IR COLtrap II}, the $\nu_2$ spectrum is clean and
apparently void of lines from other (hot or combination) bands.
Spectral line fitting was performed throughout using the standard software packages PGOPHER\cite{pgopher} and Pickett's SPFIT/SPCAT suite.\cite{pickett_JMolSpectrosc_148_371_1991}.
Molecular parameters obtained from a fit to all rotational-vibrational lines 
using a standard linear molecule Hamiltonian
are summarized in Table \ref{tbl:Fitparameter_COLtrapII_IR_Exp_und_calc}.
With only four parameters varied in the fit -- the ground vibrational-state constant $B_0$, the rotation-vibration interaction constant $\alpha_2$, a common centrifugal distortion constant $D$ for both vibrational states and the
vibrational band center $\tilde{\nu}_2$ -- the data are reproduced to within an $rms$ of 2$\times$10$^{-4}$\,\wn . 
Very good agreement between the experimental (column 'IR' in Table \ref{tbl:Fitparameter_COLtrapII_IR_Exp_und_calc}) and the calculated
rotational constants (column 'Calc') is observed with deviations not exceeding 3.7\,MHz (0.08\%). The difference between the experimental and calculated vibrational wavenumber is very small also, 0.3\,\wn\ or 0.02\%.
The agreement between the experimental ground state rotational constant and 
the corresponding best estimate value (column `BE') is even better ($<$1\,MHz, 
0.02\%). While for scaling purposes to arrive at a \nccop\ BE value of $B_0$  structurally very closely related cyanogen, NCCN, was used initially (Table~\ref{tbl:Fitparameter_COLtrapII_IR_Exp_und_calc}),
the approach has also been tested on an extended sample of ten isoelectronic (and hence potentially applicable) calibrators known from previous experiment 
(cf., section \ref{QCCsection} ``Quantum-chemical calculations" and Supporting Information). In this effort, equilibrium rotational constants $B_e$ were calculated at the ae-CCSD(T)/cc-pwCVQZ level and, 
for practical reasons, zero-point vibrational contributions
$\Delta B_0$ were computed at the more affordable fc-CCSD(T)/ANO1 level of theory.
Scaling factors derived from the ratios
$B_{0,exp}/B_{0,calc}$ in this fashion are very similar ($\sim$1.001) and invariably result in
BE values that represent a significant improvement over the calculated $B_0$
of \nccop\ alone ($|B_{\rm 0,BE}-B_{0,IR}|\le 1$\,MHz, see Supporting Information). 
This finding highlights that empirical scaling may be a worthwhile and potentially very inexpensive approach for effectively improving computational results as long as at least one isoelectronic and isostructural molecule is known from previous high-resolution study.

\subsection{\nccop\ pure rotational spectrum}
\label{sec_rot}

\begin{figure}[h!]
\centering
  \includegraphics[width=12cm]{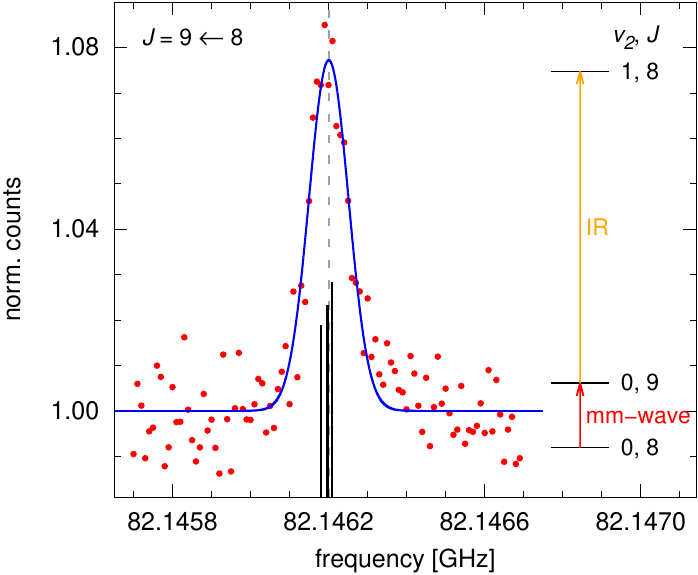}
  \caption{Pure rotational transition $J = 9 \leftarrow 8$ in the ground vibrational state of \nccop, 
  recorded using a \ac{LO} \ac{IR}/\ac{mmw} double resonance scheme. For this recording, the IR laser was stabilized on the $P(9)$ 
   transition of the $\nu_2$ vibrational mode (orange arrow) while tuning the mmw source in steps of 10\,kHz around the predicted frequency of 82.146\,GHz. 
   Here,
  resonant mmw excitation (red arrow) results in an increase of the LO signal of about 8\,\%. Normalized counts (red dots) are
  determined from the ratio of the ion counts in the mmw search window and the counts determined at an off-resonant reference frequency. Calculated quadrupole hyperfine structure from nitrogen is indicated as black sticks.
 }
  \label{fig:Rotational_Transition}
\end{figure}

Based on the molecular data obtained in the high-resolution study of the $\nu_2$ vibrational mode,
the pure rotational spectrum of \nccop\ was finally studied
using a \ac{LO}-\ac{IR}/\ac{mmw} double resonance approach
used recently with very good success in
studies of other molecular ions
\cite{asvany_hc3op_PCCP_2023,gupta_FD_245_298_2023,silva_AAA_676_L1_2023,silva_MolPhys_122_e2296613_2024,Silva_JCP_160_071101_2024,baddeliyanage_JMS_407_111978_2025}.
For this purpose, the IR laser is constantly kept fixed on the center frequency of a ro-vibrational transition 
while the millimeter-wave source is tuned in a region centered at the predicted rotational transition frequency. 
By changing the population of the lower level of the infrared transition through resonant rotational excitation, a change in the \ac{LOS} signal may be detected.
An example of this approach is given in Fig.~\ref{fig:Rotational_Transition} that shows the $J = 9 \leftarrow 8$
rotational transition of \nccop , detected through monitoring the $P(9)$ LO signal and tuning the millimeter-wave source in the region between 82145.7\,MHz and 82146.7\,MHz. The counts (red dots) were determined by
averaging the counts of several up and down scans.
Overall, using this DR technique a total of sixteen rotational transitions of \nccop\ were measured covering rotational angular momentum quantum numbers $8\le J'' \le 26$, except  $J''=15$ and 16 (Table \ref{tab:rot} and Fig. S7).
The experimental data are well described by a Gaussian line profile indicating a thermal ensemble. From a fit of the rotational transition
frequencies, the $B_0$ and $D_0$ ground state rotational parameters have been determined to much
higher precision than from the high-resolution infrared data of the $\nu_2$ mode (Table \ref{tbl:Fitparameter_COLtrapII_IR_Exp_und_calc}).
The mmw fit $rms$ obtained is 8\,kHz and hence only a very small fraction of the 
line widths of a few hundred kHz. Since the line widths show a significant
dependence on the mmw-power used, indicating power broadening effects, the experimental conditions were chosen to achieve an optimal balance between intensity and width.
No hyperfine structure from the presence of the \ce{^14N} nucleus ($I=1$) was spectroscopically resolved
in the pure rotational spectra (see Fig.~\ref{fig:Rotational_Transition}).
However, an accurate quadrupole coupling parameter is obtained from calculations; at the ae-CCSD(T)/cc-pwCVQZ level, this amounts to $eQq=-5.880$\,MHz.
It turns out that this value is very similar to the corresponding one found for the terminal
nitrogen nucleus in the \ce{NCCNH+} molecular ion ($-5.480(3)$\,MHz) with Fourier transform microwave spectroscopy,\cite{gottlieb_JCP_113_1910_2000} an indication that the nitrogen nuclei share a very similar electronic environment in both ions.
Empirical scaling using an ae-CCSD(T)/cc-pwCVQZ structural calculation
of \ce{NCCNH+} results in a corrections factor $eQq_{exp}/eQq_{calc}=-5.480/-5.532$ close to unity and hence a very small impact on
the predicted $eQq_{calc}$ of \nccop . The BE value obtained in this fashion is $-5.827$\,MHz and should permit very accurate predictions of the quadrupole hyperfine structure in the rotational spectrum of \nccop , in particular for 
low-$J$ transitions where the strong $\Delta F=+1$ components might be resolvable. 

\begin{table}[h!]
\centering
\caption{Rotational transition frequencies of \nccop\ in the ground vibrational state (in MHz) and fit residuals $o-c$ (kHz). 
}
\label{tab:rot} 
\begin{tabular}{r@{}lr@{}lr@{}l}
\hline
$J'		$~ & $ \leftarrow	J''$		&	\multicolumn{2}{c}{Experimental}	&	\multicolumn{2}{c}{$o-c$} \\
\hline														
$	9	$~ & $ \leftarrow	8	$	&	82146	&	.1978	& $-$0	&	.2	\\
$	10	$~ & $ \leftarrow	9	$	&	91273	&	.0957	&	 5	&	.7	\\
$	11	$~ & $ \leftarrow	10	$	&	100399	&	.8551	&  	19	&	.3	\\
$	12	$~ & $ \leftarrow	11	$	&	109526	&	.4239	& 	 3	&	.3	\\
$	13	$~ & $ \leftarrow	12	$	&	118652	&	.8269	&  $-$3	&	.0	\\
$	14	$~ & $ \leftarrow	13	$	&	127779	&	.0479	&  $-$1	&	.2	\\
$	18	$~ & $ \leftarrow	17	$	&	164281	&	.7258	&  $-$5	&	.3	\\
$	19	$~ & $ \leftarrow	18	$	&	173406	&	.7764	&  $-$3	&	.5	\\
$	20	$~ & $ \leftarrow	19	$	&	182531	&	.5372	& $-$13	&	.5	\\
$	21	$~ & $ \leftarrow	20	$	&	191656	&	.0360	&	  7	&	.1	\\
$	22	$~ & $ \leftarrow	21	$	&	200780	&	.1894	& $-$10	&	.5	\\
$	23	$~ & $ \leftarrow	22	$	&	209904	&	.0435	&  $-$5	&	.6	\\
$	24	$~ & $ \leftarrow	23	$	&	219027	&	.5667	&	  5	&	.0	\\
$	25	$~ & $ \leftarrow	24	$	&	228150	&	.7249	&	  1	&	.6	\\
$	26	$~ & $ \leftarrow	25	$	&	237273	&	.5280	&	  8	&	.8	\\
$	27	$~ & $ \leftarrow	26	$	&	246395	&	.9347	&	  0	&	.1	\\
\hline														
\end{tabular}
\end{table}

\section{Conclusions and Outlook} 

This paper reports on the first spectroscopic detection and characterization of the linear \nccop~ion, achieved by employing highly sensitive action spectroscopic methods in cryogenic ion trap apparatus. 
Initial spectroscopic detection was accomplished using \ac{IRPD} spectroscopy of the weakly bound \nccop --Ne 
cluster followed-up upon by high-resolution \ac{IR}-\ac{LOS} of the $\nu_2$ fundamental of bare \nccop\
and concluded with \ac{LO} \ac{IR}/\ac{mmw} double resonance spectroscopy of 16 pure rotational transitions
between 82 and 247\,GHz.
Based on the experimental and calculated molecular parameters, the rotational spectrum of \nccop\ in its ground vibrational state can now be predicted accurately from the microwave well into the submillimeter-wave range, a prerequisite
for dedicated and sensitive radio astronomical searches.
Although \nccop\ is not currently included in standard astrochemical models,\cite{Wakelam2012,Millar2024} its close structural and compositional similarity to known astronomical species makes it an attractive candidate for future astronomical searches.
Frequency predictions will be provided through a corresponding entry in the Cologne Database for Molecular spectroscopy
\cite{mueller_cdms,endres_cdms_2016}.

Besides high-resolution characterization of other vibrational bands of \nccop\ such as the strong $\nu_1$ mode at 2340\,\wn , future extension towards spectroscopic characterization of other [2C,N,O\ce{]+} structural isomers seems promising. Energetically, ground state \nccop\ is followed by
linear \ce{CNCO+} calculated to lie only 11\,kcal/mol above\cite{chi_JMStTheo_763_91_2006}. Some evidence for the production of \ce{CNCO+} has been obtained already previously through
mass spectrometry\cite{mcgibbon_IJMSIP_121_R11_1992}. By comparison,
in the isoelectronic [2C,2N] system, at least three structural isomers are known spectroscopically, two by observations at high-resolution in the gas phase, 
cyanogen, NCCN,\cite{maki_JMS_269_166_2011} and isocyanogen, CNCN\cite{gerry_JMS_140_147_1990} (located +26\,kcal/mol above NCCN\cite{ding_JCP_108_2024_1998}) and a third one, diisocyanogen, CNNC, trapped in a matrix of solid Ar at low temperatures\cite{maier_AngChemIntEd_31_1218_1992} (+74\,kcal/mol). 

Laboratory studies of new thioacylium species, $R-$\ce{CS+}, in which the oxygen atom of acylium ions
is replaced with isovalent sulfur would be another interesting area of research.
No [2C,N,S\ce{]+} species have yet been characterized spectroscopcially,
but calculations indicate linear \ce{NCCS+} to represent the global minimum arrangement\cite{chi_JMStTheo_766_165_2006}.
In the present study, exploratory anharmonic fc-CCSD(T)/ANO2 calculations of 
\ce{NCCS+} indicate the vibrational spectrum to be dominated by the strong C-N and C-S stretching fundamental bands located at 2175\,\wn\ and 1630\,\wn , respectively, and hence promising targets for future infrared studies. Also, at the ae-CCSD(T)/cc-pwCVQZ level, \ce{NCCS+} is calculated to be very polar, with a center-of-mass dipole moment of 3.77 D and a rotational constant of about 2.8 GHz, making it an attractive candidate for studies of its pure rotational spectrum in the microwave region.
Similar to the scaling approach performed for the \ce{HC3S+} molecular ion,\cite{thorwirth_MolPhys_118_e1776409_2020}
using isoelectronic NCCP\cite{bizzocchi_JMS_110_205_2001} for calibration purposes, should provide a rotational constant of very high accuracy, possibly as good as to within 1\,MHz (cf. Refs.
\citenum{thorwirth_MolPhys_118_e1776409_2020,cernicharo_AAA_646_L3_2021}).
It might be worthwhile to mention that the Fourier transform microwave spectrum of the kinked NCCS radical has been known already for 20 years \cite{mccarthy_ApJSS_144_287_2003}.

In summary, the present work demonstrates the power of action spectroscopy in cold ion traps where different variants are performed in a sequential fashion: Initially,
the vibrational
bands of target species are indentified in low-resolution studies followed by investigations at high resolution unfolding
their rotational substructures. Finally, pure rotational spectra may be recorded at even much higher resolution as needed
for subsequent radio astronomical searches. The fast scanning speed of leak-out spectroscopy is unprecedented and will greatly support the study of many
new ions of astrophysical interest in the future.


\section*{Conflicts of interest}
There are no conflicts of interest to declare.


\begin{acknowledgement}

We are grateful to Prof.\;Dr.\;Britta Redlich for a generous donation of \ac{FELIX} director's beamtime for the \ac{IRPD} study, Radboud University, the Nederlandse
Organisatie voor Wetenschappelijk Onderzoek (NWO) 
as well as the \ac{FELIX} staff for support.
This work  has been  supported by an
ERC advanced grant (MissIons: 101020583),
by the Deutsche Forschungsgemeinschaft (DFG) via the Collaborative Research Centre 1601 (project ID: 500700252, sub-project B8 and C4)
and the Ger\"atezentrum ``Cologne Center for Terahertz Spectroscopy" (DFG SCHL 341/15-1).
We thank Florian Pirlet for technical support and the Regional Computing Center of the Universit\"at zu K\"oln (RRZK) for providing computing time on the DFG-funded high-performance computing system CHEOPS.
The authors gratefully acknowledge the work carried out in recent years by the electrical and mechanical workshops of the I. Physikalisches Institut, without whose support the construction of the new ion trap apparatus COLtrap~II would not have been possible.
\end{acknowledgement}

\begin{suppinfo}

The Supporting Information is available free of charge
providing further details on quantum-chemical calculations, the ion temperature, mass spectra, enlarged ro-vibrational spectra, IR linelist, and spectra of pure rotational transitions.

\end{suppinfo}

\bibliography{LIRTRAP,sthorwirth_bibdesk_NCCO+_revision}

\end{document}